\documentclass[twocolumn,aps,prl,showpacs]{revtex4}
\usepackage{amsfonts}
\usepackage{amsmath}
\usepackage{graphicx}

\input{tcilatex}

\begin{document}

\title{Nonlinear Inequalities and Entropy-Concurrence Plane}
\author{Fabio Antonio Bovino}
\email{fabio.bovino@elsag.it}
\affiliation{Elsag spa, Via Puccini 2-16154 Genova, Italy}
\date{\today }

\begin{abstract}
Nonlinear inequalities based on the quadratic Renyi entropy for mixed
two-qubit states are characterized on the Entropy-Concurrence plane. This
class of inequalities is stronger than Clauser-Horne-Shimony-Holt (CHSH)
inequalities and, in particular, are violated "in toto" by the set of Type I
Maximally-Entangled-Mixture States (MEMS I).
\end{abstract}

\pacs{ }
\maketitle







\section{Introduction}

Entanglement, \textquotedblleft \emph{the} characteristic trait of quantum
mechanics\textquotedblright ~\cite{Schrodinger35}, has been identified as a
fundamental physical resource for quantum computation and information, and
its quantification and detection have been the subject of considerable
research. However, despite a remarkable progress in the field, the so-called 
\emph{separability problem}, the question whether a state $\varrho $ is
entangled or not, has not yet a general answer.

More precisely, a quantum state described by density matrix $\varrho $ of a
system composed of two subsystems of dimension $N$ and $M$, respectively, is
called entangled \cite{Werner89} iff it cannot be written as a separable
state of the form 
\begin{equation}
\sigma =\tsum\limits_{k}p_{k}\left\vert \psi _{k}\right\rangle \left\langle
\psi _{k}\right\vert \otimes \left\vert \phi _{k}\right\rangle \left\langle
\phi _{k}\right\vert,
\end{equation}%
where $p_{k}\geq 0$ and $\tsum p_{k}=1$.

Currently the most important criterion for deciding whether a given state is
entangled or not is related to the semidefinite positivity of the partial
transpose $\varrho ^{T_{A}}$: separable states have a positive semidefinite
partial transpose PPT, hence all non-PPT states are entangled. For systems
with $2\times 2$ and $2\times 3$ dimensional Hilbert spaces the
PPT-criterion also turned out to be sufficient \cite{Horodecki96}, but for
higher dimensional systems there exist PPT entangled states.

Further, a complete characterization of separable states exists based on "%
\emph{entanglement witness}". Briefly speaking, entanglement witnesses are
operators that are designed directly for distinguishing between separable
and entangled states\cite{Peres96,Horodecki96,Terhal00}. A Hermitian
operator $W$ is called an entanglement witness if it has a positive
expectation value with respect to all separable states, $Tr\left( W\sigma
\right) \geq 0.$ The negative expectation value is hence a signature of
entanglement, and a state with $Tr\left( W\varrho \right) <0$ is said to be
detected by the witness. The latter condition offers the possibility of
experimental detection of entanglement via the measurement of $W$, an
observable which \textquotedblleft \emph{witnesses}\textquotedblright\ the
quantum correlations in $\varrho $.

Historically, a violation of Bell's inequalities~\cite{Bell64,CHSH69,FC72}
provided the first test for entanglement. Bell's inequalities were
originally designed to prove that quantum mechanics is incompatible with
Einstein, Podolsky, and Rosen (EPR) local realistic view of the world~\cite%
{EPR35} but, within quantum mechanics, they can be also regarded as
non-optimal linear witness operators.

Geometrically, separable states form a convex set in the space of all
density matrices of a given system and one might expect that special types
of nonlinear witnesses can approximate the convex set of the separable
states better than linear ones. In particular, following Schroedinger
remarks on relations between the information content of the total system and
its subsystems, some separability criteria in terms of entropic uncertainty
relations were derived.

Classically, if a system is formed by different subsystems, complete
knowledge of the whole system implies that the sum of the information of the
subsystems makes up the complete information for the whole system. The
Shannon entropy $H\left( X\right) $ of a single random variable is never
larger than the Shannon entropy of two random variables, that is: $H\left(
X,Y\right) \geq H\left( X\right) ,H\left( Y\right) $. In the quantum world,
there exist states of composite systems for which we might have the complete
information, while our knowledge about the subsystems might be very poor or
null. The canonical example is given by a pair of qubits A and B prepared in
the maximally entangled state $\left( \left\vert 00\right\rangle +\left\vert
11\right\rangle \right) /\sqrt{2}$. The von Neumann entropy $S(A)$ of qubit
A is equal to $1$, compared with a von Neumann entropy $S\left( A,B\right) $
of $0$ for the joint system. It has been shown \cite%
{Nielsen98,CerfAdami99,Horodecki98} that for separable states the relation%
\begin{equation}
S(A,B)\geq S\left( A\right) ,S\left( B\right) ,  \label{Neumann}
\end{equation}%
holds as a consequence of the its concavity \cite{Lieb73} but,
unfortunately, the inequalities (\ref{Neumann}) are not sufficient to
characterize separability.

The idea to use higher order (nonlinear) entropic inequalities as
separability-vs-entanglement criteria for mixed states born when Cerf and
Adami \cite{CerfAdami97} and the Horodecki family \cite%
{Horodecki96,RHorodecki96} recognized that conditional R\.{e}nyi entropies
are non-negative for separable states and it was recently proposed by
several groups \cite{Abe01,Abe99,Tsallis01,Barranco99,Rajagopal01,Alcaraz01}%
, in the form of conditional Tsallis entropies. These entropic inequalities
are satisfied by all separable states and are known to be stronger than all
Bell-CHSH inequalities.

Recently Derkacz and Jac\'{o}bczyk \cite{Derkacz04,Derkacz05} studied the
relationship between entanglement, as measured by concurrence $C\left(
\varrho \right) $, mixedness, as measured by linear entropy $S_{L}\left(
\varrho \right) $, and Bell-CHSH violation. These authors showed that the
subset $\Lambda $ on the $\left( C,S_{L}\right) $ plane, previously
investigate by Munro et al. \cite{Munro01}, is the sum of disjoint subsets $%
\Lambda _{V}$, $\Lambda _{NV}$ and $\Lambda _{0}$, with the following
properties: states belonging to $\Lambda _{V}$ violate CHSH inequalities,
states belonging to $\Lambda _{NV}$ satisfy CHSH inequalities, states from $%
\Lambda _{0}$, different but with the same entropy and concurrence, can
violate or satisfy CHSH inequalities.

Following Derkacz and Jac\'{o}bczyk, in this paper the relationship between
two-qubit states entanglement and the violation of entropic inequalities on
the $\left( C,S_{L}\right) $ plane is investigated.

\section{Non-linear Entropies}

The quantum R\.{e}nyi entropy depending on the entropic parameter $\alpha
\in 
\mathbb{R}
$ is given by 
\begin{equation}
S_{\alpha }\left( \varrho \right) =\frac{\log \text{ }Tr\left( \varrho
^{\alpha }\right) }{1-\alpha },
\end{equation}%
where $S_{0}$, $S_{1}$, $S_{\infty }$ reduce to the logarithm of the rank,
the von Neumann entropy and the negative logarithm of the operator norm,
respectively. The conditional R\.{e}nyi entropy reads%
\begin{equation}
S_{\alpha }\left( B|A;\text{ }\varrho \right) :=S_{\alpha }\left( \varrho
\right) -S_{\alpha }\left( \varrho _{A}\right) .
\end{equation}%
The Tsallis entropy, given by%
\begin{equation}
T_{\alpha }\left( \varrho \right) :=\frac{1-Tr\left( \varrho ^{\alpha
}\right) }{1-\alpha },
\end{equation}%
is non-negative, concave (convex) for $\alpha >0$ ($\alpha <0$) and reduces
the von Neumann entropy in the limit $\alpha \rightarrow 1$. The conditional
Tsallis entropy reads%
\begin{equation}
T_{\alpha }\left( B|A;\text{ }\varrho \right) =\frac{Tr\left( \varrho
_{A}^{\alpha }\right) -Tr\left( \varrho ^{\alpha }\right) }{\left( 1-\alpha
\right) Tr\left( \varrho _{A}^{\alpha }\right) }.
\end{equation}%
Concerning positivity, however, the two conditional entropies are
equivalent, i.e:%
\begin{equation}
T_{\alpha }\left( B|A;\text{ }\varrho \right) \geq 0\text{ and }S_{\alpha
}\left( B|A;\text{ }\varrho \right) \geq 0,
\end{equation}%
which is equivalent to%
\begin{eqnarray}
Tr\left( \varrho _{A}^{\alpha }\right) &\geq &Tr\left( \varrho ^{\alpha
}\right) \text{ for }\alpha >1, \\
Tr\left( \varrho _{A}^{\alpha }\right) &\leq &Tr\left( \varrho ^{\alpha
}\right) \text{ for }0\leq \alpha <1.  \notag
\end{eqnarray}%
The conditional Tsallis/R\.{e}nyi entropies, involving higher power ($\alpha
>1$) of density matrix $\varrho $, provide a more stringent criterion for
separability\cite{Wolf02}.

\section{Entropic inequalities and entropy-concurrence plane.}

The aim of this section is to obtain the subset of the entanglement-mixdness
plane corresponding to violation of quadratic entropic inequalities ($\alpha
=2$). In this case it is possible to extract a nonlocal and nonlinear
quantity, namely, the Renyi entropy, from local measurements on two pairs of
polarization-entangled photons as showed in \cite{Fabio05}.

The entanglement can be quantified by the quantity $C\left( \varrho \right) $
which is known in literature as \emph{concurrence. }Wooters has derived an
analytic formula for the concurrence of two-qubit states:%
\begin{equation}
C\left( \varrho \right) =2\max \{\lambda _{j}\}-\tsum\limits_{j}\lambda _{j},
\end{equation}%
where $\lambda _{j}$ are the square roots of eigenvalues of the matrix $%
\tilde{\varrho}=\varrho \left( \sigma _{y}\otimes \sigma _{y}\right) \varrho
^{\ast }\left( \sigma _{y}\otimes \sigma _{y}\right) $ and $\varrho ^{\ast }$
denotes the complex conjugate of density operator $\varrho $. The mixedness
measure is the so-called \emph{linear entropy }and is based on the purity of
a state $P=Tr\left( \varrho ^{2}\right) $. The linear entropy $S_{L}$ for $%
C^{2}\otimes C^{2}$ systems is defined via%
\begin{equation}
S_{L}=\frac{4}{3}\left[ 1-Tr\left( \varrho ^{2}\right) \right] ,
\end{equation}%
and ranges from $0$ to $1$ (for a maximally mixed state).

In the entanglement-mixedness or, in this case, concurrence-entropy plane,
we can start considering the class the class $\mathcal{E}_{0}$ of states%
\begin{equation}
\varrho =%
\begin{pmatrix}
0 & 0 & 0 & 0 \\ 
0 & a & \frac{1}{2}ce^{i\vartheta } & 0 \\ 
0 & \frac{1}{2}ce^{-i\vartheta } & b & 0 \\ 
0 & 0 & 0 & 1-a-b%
\end{pmatrix}%
,  \label{density}
\end{equation}%
where%
\begin{eqnarray}
c &\in &\left[ 0,1\right] ,\text{ }a,b\geq 0,\text{ }\vartheta \in \left[
0,2\pi \right] ,  \notag \\
ab &\geq &\frac{c^{2}}{4}\text{ and }a+b\leq 1,  \label{conditions}
\end{eqnarray}%
from the positive definiteness of $\varrho $. For the class of states $%
\mathcal{E}_{0}$, the normalized linear entropy reads%
\begin{equation}
S_{L}=\frac{4}{3}\left( 1-a^{2}-b^{2}-\left( 1-\left( a+b\right) \right)
^{2}-\frac{c^{2}}{2}\right) ,  \label{linearentropy}
\end{equation}%
and the concurrence is given by%
\begin{equation}
C\left( \varrho \right) =c.
\end{equation}%
The boundary value of (\ref{linearentropy}) for fixed $c$, $a$ and $b$, such
that conditions (\ref{conditions}) are satisfied, is given by%
\begin{equation}
S_{L\max 1}\left( c\right) =\frac{8}{3}c\left( 1-c\right) ,
\label{frontier1}
\end{equation}%
for $c\in \left[ \frac{2}{3},1\right] $ and $a=b=\frac{c}{2},$%
\begin{equation}
S_{L\max 2}\left( c\right) =\frac{8}{9}-\frac{2}{3}c^{2},  \label{frontier2}
\end{equation}%
for $c\in \left( 0,\frac{2}{3}\right) $ and $a=b=\frac{1}{3}$. Then $%
S_{L\max }\left( c\right) $ is reached by the so-called \emph{Maximally
Entangled Mixed States }(MEMS's). These are the states which maximize the
entanglement degree for a given value of the linear entropy (purity). In
particular we can distinguish two families of these states, I and II,
defined as%
\begin{equation}
\varrho _{1}\left( c\right) =%
\begin{pmatrix}
0 & 0 & 0 & 0 \\ 
0 & \frac{c}{2} & \frac{1}{2}ce^{i\vartheta } & 0 \\ 
0 & \frac{1}{2}ce^{-i\vartheta } & \frac{c}{2} & 0 \\ 
0 & 0 & 0 & 1-c%
\end{pmatrix}%
,\text{ MEMS I}  \label{MEMS I}
\end{equation}%
\begin{equation}
\varrho _{2}\left( c\right) =%
\begin{pmatrix}
0 & 0 & 0 & 0 \\ 
0 & \frac{1}{3} & \frac{1}{2}ce^{i\vartheta } & 0 \\ 
0 & \frac{1}{2}ce^{-i\vartheta } & \frac{1}{3} & 0 \\ 
0 & 0 & 0 & \frac{1}{3}%
\end{pmatrix}%
,\text{ MEMS II}  \label{MEMS II}
\end{equation}%
Now let us consider the structure of the set $\Lambda _{\mathcal{E}_{0}}$
defined by the frontiers (\ref{frontier1}) and (\ref{frontier2}). 
\begin{figure}[t]
\centerline{\includegraphics[width=7.6cm]{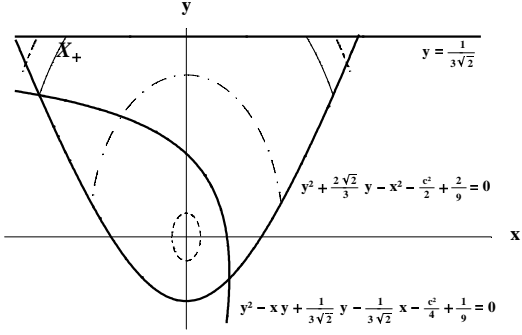}}
\caption{The figure shows the region of plane $\protect\varrho \in \mathcal{E%
}_{0}$ defined by the point $\left( x,y\right) \in X_{+}$ bounded by the
hyperbola $y^{2}+\frac{2\protect\sqrt{2}}{3}y-x^{2}-\frac{c^{2}}{2}+\frac{2}{%
9}=0$ and by the straight line $y=\frac{1}{3\protect\sqrt{2}}$. For fixed
concurrence $c$ the intersection of the level set of the function $S_{L}$
(represented by dotted lines) with $X_{+}$ can lie below or above the curve
representing the inequality bound (\protect\ref{eineq}), or can intersect
this line, depending on the value $s$.}
\label{fig1}
\end{figure}
\begin{theorem}
Entropic inequalities disjoin the set $\Lambda _{\mathcal{E}_{0}}$ in a sum
subsets $\Lambda _{V_E}$, $\Lambda _{0_E}$ and $\Lambda _{NV_E}$:
\end{theorem}

\begin{enumerate}
\item If $\left( s,c\right) \in \Lambda _{V\_E}$, then every state $\varrho
\in \mathcal{E}_{0}$ such that $S_{L}\left( \varrho \right) =s$ and $C\left(
\varrho \right) =c$ violates entropic inequalities.

\item If $\left( s,c\right) \in \Lambda _{0\_E}$, then there exist states $%
\varrho _{1},\varrho _{2}\in \mathcal{E}_{0}$ such that $S_{L}\left( \varrho
_{1}\right) =S_{L}\left( \varrho _{2}\right) =s$ and $C\left( \varrho
_{1}\right) =C\left( \varrho _{2}\right) =c,$ but $\varrho _{1}$ violates
entropic inequalities, while $\varrho _{2}$ does not violate entropic
inequalities.

\item If $\left( s,c\right) \in \Lambda _{NV\_E}$, then every state $\varrho
\in \mathcal{E}_{0}$ such that $S_{L}\left( \varrho \right) =s$ and $C\left(
\varrho \right) =c$ does not violate entropic inequalities.
\end{enumerate}

\begin{proof}
\end{proof}Following Derkacz and Jac\'{o}bczyk Let us introduce the new
variables%
\begin{equation}
x=\frac{1}{\sqrt{2}}\left( a-b\right) ,\text{ }y=\frac{1}{\sqrt{2}}\left(
a+b-\frac{2}{3}\right) ,
\end{equation}%
Each state $\varrho \in \mathcal{E}_{0}$ is now defined by the point $\left(
x,y\right) \in X_{+}$where%
\begin{equation}
X_{+}=\{\left( x,y\right) :y^{2}+\frac{2\sqrt{2}}{3}y-x^{2}-\frac{c^{2}}{2}+%
\frac{2}{9}\geq 0,\text{ }y\leq \frac{1}{3\sqrt{2}}\},
\end{equation}%
and linear entropy $S_{L}\left( \varrho \right) $ is now expressed as%
\begin{equation}
S_{L}\left( \varrho \right) =-\frac{8}{3}\left( \frac{x^{2}}{2}+\frac{3}{2}%
y^{2}+\frac{c^{2}}{2}-\frac{1}{3}\right) .
\end{equation}%
\begin{figure}[t]
\centerline{\includegraphics[width=7.6cm]{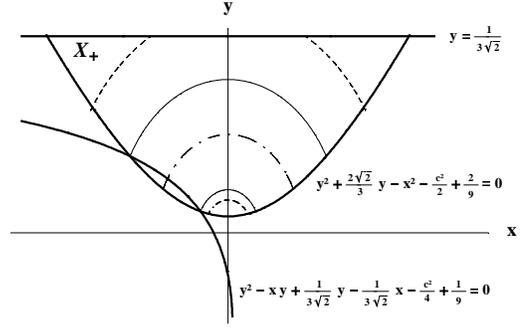}}
\caption{The curve representing entropic condition intersects the hyperbola
only in two points because they have a common asymptote ($y=x-\frac{\protect%
\sqrt{2}}{3}$). The upper one always lies in the halfplane $x<0$ and until
the intersection of the ellipse representing level set of $S_{L}$ with the
hyperbola is above this point, i.e. for $s<\frac{1}{3}\left( 1+c^{2}-\protect%
\sqrt{1-2c^{2}}\right) $, all states are VEIS. The lower point for $c>\frac{2%
}{3}$ lies in the halfplane $x<0$ and until the intersection of the ellipse
representing level set of $S_{L}$ with the hyperbola is below this point,
i.e. $s>\frac{1}{3}\left( 1+c^{2}+\protect\sqrt{1-2c^{2}}\right) $, all
states are VEIS.}
\end{figure}
Then the states with the same value $S_{L}=s$ belong to the ellipse%
\begin{equation}
\frac{x^{2}}{A^{2}}+\frac{y^{2}}{B^{2}}=1,
\end{equation}%
with%
\begin{equation}
A=\sqrt{6\left( -\frac{c^{2}}{12}-\frac{s}{8}+\frac{1}{9}\right) },\text{ }B=%
\sqrt{2\left( -\frac{c^{2}}{12}-\frac{s}{8}+\frac{1}{9}\right) }.
\end{equation}%
The entropic inequality $Tr\left( \varrho ^{2}\right) -Tr\left( \varrho
_{A}^{2}\right) \leq 0$ now reads%
\begin{equation}
y^{2}-xy+\frac{1}{3\sqrt{2}}y+\frac{1}{3\sqrt{2}}x-\frac{c^{2}}{4}+\frac{1}{9%
}\leq 0  \label{eineq}
\end{equation}%
For fixed concurrence $c$ the intersection of the level set of the function $%
S_{L}$ with $X_{+}$ can lie below or above the curve representing the
inequality bound (\ref{eineq}), or can intersect this line, depending on the
value $s$.

The ellipse can intersect the inequality function (\ref{eineq}) for $s\leq 
\frac{8}{3}c\left( 1-c^{2}\right) $ for $c>\frac{1}{2}$ and $s\leq \frac{2}{3%
}$ for $0<c\leq \frac{1}{2}$: the part of ellipse above hyperbola $y^{2}+%
\frac{2\sqrt{2}}{3}y-x^{2}-\frac{c^{2}}{2}+\frac{2}{9}=0$ represents
Violating Entropic Inequalities States (VEIS), whereas the remaining part
corresponds to states with the same $s$ and $c$, which are not VEIS (see Fig.%
\ref{fig1}).

For $s>\frac{2}{3}$ no state violates the entropic inequality. 
\begin{figure}[t]
\centerline{\includegraphics[width=7.6cm]{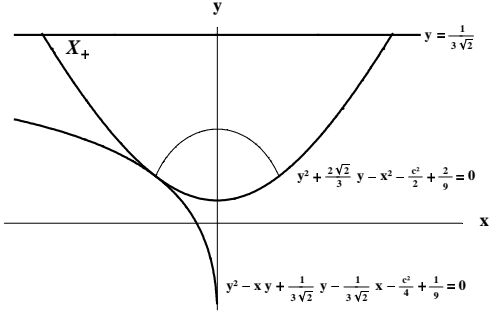}}
\caption{For $c>\frac{1}{\protect\sqrt{2}}$ the curve representing the
inequality bound has no common points with $X_{+}$ and all states are VEIS.}
\end{figure}
The curve representing entropic condition intersects the hyperbola $y^{2}+%
\frac{2\sqrt{2}}{3}y-x^{2}-\frac{c^{2}}{2}+\frac{2}{9}=0$ only in two points
because they have a common asymptote ($y=x-\frac{\sqrt{2}}{3}$). The upper
one always lies in the halfplane $x<0$ and until the intersection of the
ellipse with the hyperbola is above this point, i.e. for $s<\frac{1}{3}%
\left( 1+c^{2}-\sqrt{1-2c^{2}}\right) $, all states are VEIS. The lower
point for $c>\frac{2}{3}$ lies in the halfplane $x<0$ and until the
intersection of the ellipse with the hyperbola is below this point, i.e. $s>%
\frac{1}{3}\left( 1+c^{2}+\sqrt{1-2c^{2}}\right) $, all states are VEIS.

For $c>\frac{1}{\sqrt{2}}$ the curve representing the inequality bound has
no common points with $X_{+}$ and all states are VEIS. 
\begin{figure}[b]
\centerline{\includegraphics[width=7.6cm]{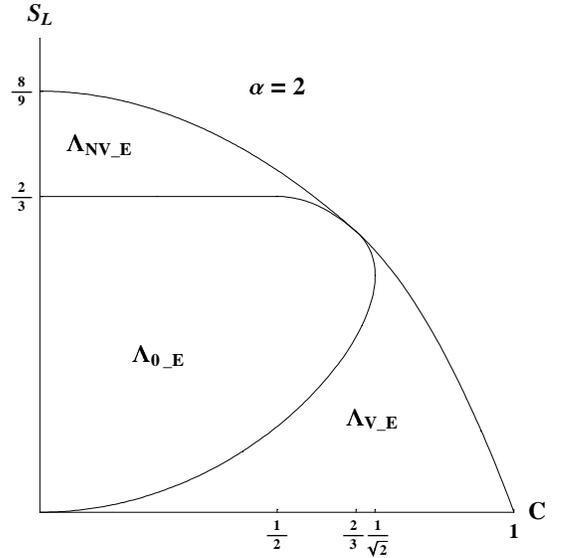}}
\caption{The Figure represents the structure of the set $\Lambda _{\mathcal{E%
}_{0}}$ defined by the frontiers (\protect\ref{frontier1}) and (\protect\ref%
{frontier2}). If $\left( s,c\right) \in \Lambda _{V\_E}$, then every state $%
\protect\varrho \in \mathcal{E}_{0}$ such that $S_{L}\left( \protect\varrho %
\right) =s$ and $C\left( \protect\varrho \right) =c$ violates entropic
inequalities. If $\left( s,c\right) \in \Lambda _{0\_E}$, then there exist
states $\protect\varrho _{1},\protect\varrho _{2}\in \mathcal{E}_{0}$ such
that $S_{L}\left( \protect\varrho _{1}\right) =S_{L}\left( \protect\varrho %
_{2}\right) =s$ and $C\left( \protect\varrho _{1}\right) =C\left( \protect%
\varrho _{2}\right) =c$ and $\protect\varrho _{1}$ violates entropic
inequalities, but $\protect\varrho _{2}$ does not violate entropic
inequalities. If $\left( s,c\right) \in \Lambda _{NV\_E}$, then every state $%
\protect\varrho \in \mathcal{E}_{0}$ such that $S_{L}\left( \protect\varrho %
\right) =s$ and $C\left( \protect\varrho \right) =c$ does not violates
entropic inequalities.}
\end{figure}
These conditions define the subsets $\Lambda _{NV\_E}$, $\Lambda _{V\_E}$
and $\Lambda _{0\_E}$.%
\begin{eqnarray}
\Lambda _{NV\_E} &=&\{\left( s,c\right) :0<c<\frac{1}{2},\frac{2}{3}<s\leq
S_{L2}\left( c\right) \}  \notag \\
\cup \{\left( s,c\right) &:&\frac{1}{2}\leq c<\frac{2}{3},S_{L1}\left(
c\right) \leq s\leq S_{L2}\left( c\right) \},  \notag \\
\Lambda _{V\_E} &=&\{\left( s,c\right) :0<c<\frac{1}{\sqrt{2}},0\leq
s<S_{L-}\left( c\right) \}  \notag \\
\cup \{\left( s,c\right) &:&\frac{2}{3}\leq c\leq \frac{1}{\sqrt{2}}%
,S_{L+}\left( c\right) <s\leq S_{L1}\left( c\right) \}  \notag \\
\cup \{\left( s,c\right) &:&c>\frac{1}{\sqrt{2}},0\leq s\leq S_{L1}\left(
c\right) \},  \notag \\
\Lambda _{0\_E} &=&\{\left( s,c\right) :\notin \Lambda _{V},\Lambda _{NV}\},
\label{subsets}
\end{eqnarray}%
where%
\begin{eqnarray}
S_{L1}\left( c\right) &=&S_{L\max 1}\left( c\right) ,  \notag \\
S_{L2}\left( c\right) &=&S_{L\max 2}\left( c\right) ,  \notag \\
S_{L+}\left( c\right) &=&\frac{1}{3}\left( 1+c^{2}+\sqrt{1-2c^{2}}\right) , 
\notag \\
S_{L-}\left( c\right) &=&\frac{1}{3}\left( 1+c^{2}-\sqrt{1-2c^{2}}\right) .
\label{curves}
\end{eqnarray}%
Fig.\ref{CHSHtoo} represents the bounds fixed by the Entropic Inequality in
respect to CHSH Inequality. We can see that the region of $(C,S_{L})$ plane
where entanglement is detected is larger, and there is a reduction of the
region where the entanglement is not detected. 
\begin{figure}[b]
\centerline{\includegraphics[width=7.6cm]{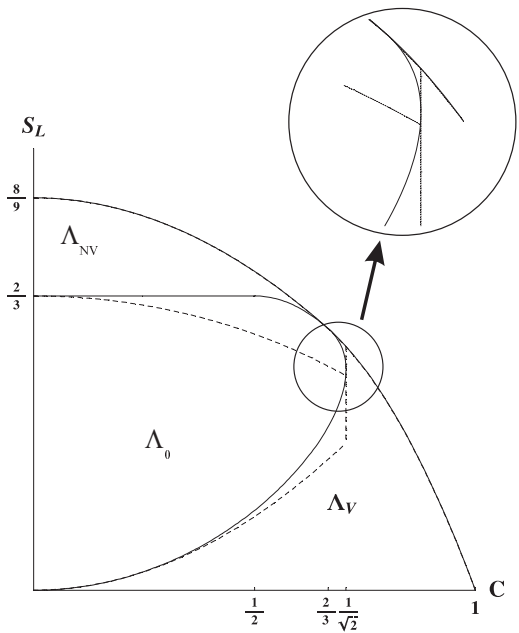}}
\caption{The Figure shows the bounds found from the Nonlinear Inequality
(continous line) and CHSH inequality (dotted line) on Entropy-Concurrrence
plane. It is possible to appreciate the larger region where entanglement is
detected. The branch of MEMs I for $2/3\leq c\leq 1/\protect\sqrt{2}$,
violates "\emph{in toto}" the nonlinear inequality, as showed in the inset.}
\label{CHSHtoo}
\end{figure}
We can quantify these results in terms of relative area of the different
subsets $\Lambda _{V\_E}$, $\Lambda _{0\_E}$ and $\Lambda _{NV\_E}$ in
respect to the total area of the total set $\Lambda $ corresponding to
physical states and we can compare the results with CHSH case:%
\begin{eqnarray}
\Lambda _{V\_E} &\simeq &28.390\%\text{ \ \ }\Lambda _{V\_CHSH}\simeq
26.577\%  \notag \\
\Lambda _{0\_E} &\simeq &58.155\%\text{ \ \ \ }\Lambda _{0\_CHSH}\simeq
54.788\%  \notag \\
\Lambda _{NV\_E} &\simeq &13.455\%\text{ \ \ \ }\Lambda _{NV\_CHSH}\simeq
18.635\%
\end{eqnarray}%
It is possible to appreciate the larger region where entanglement is
detected by nonlinear inequality in respect to CHSH ones. Moreover CHSH
inequality does not detect the branch of MEMs I for $2/3\leq c\leq 1/\sqrt{2}
$, while nonlinear inequality are violated by these states "\emph{in toto}"
as showed in Fig.\ref{CHSHtoo}.

\begin{figure}[t]
\centerline{\includegraphics[width=7.6cm]{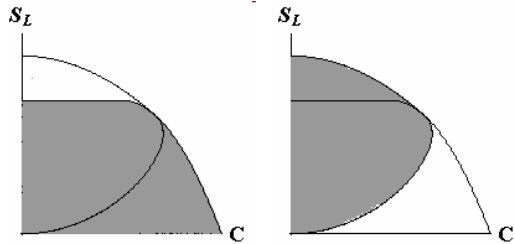}}
\caption{The numerical analysis of the set $\Lambda _{\mathcal{E}_{1}}$
shows that its structure is the same of the previous set of density matrices 
$\Lambda _{\mathcal{E}_{0}}$, i.e. the bounds of the the three regions $%
\Lambda _{V\_E}$, $\Lambda _{0\_E}$ and $\Lambda _{NV\_E}$ remain unchanged.
The picture on the left shows the states which violate the non-linear
inequality, the picture on the right the states which satisfy the
inequality. }
\label{fig6}
\end{figure}

Let us extend this results to the larger class $\mathcal{E}_{1}$of states of
the form%
\begin{equation}
\varrho =%
\begin{pmatrix}
f & 0 & 0 & \frac{1}{2}de^{i\phi } \\ 
0 & a & \frac{1}{2}ce^{i\vartheta } & 0 \\ 
0 & \frac{1}{2}ce^{-i\vartheta } & b & 0 \\ 
\frac{1}{2}de^{-i\phi } & 0 & 0 & 1-a-b-f%
\end{pmatrix}%
,
\end{equation}%
For these states the normalized linear entropy reads%
\begin{equation}
S_{L}=\frac{4}{3}\left( 1-a^{2}-b^{2}-\frac{c^{2}}{2}-\frac{d^{2}}{2}%
-f^{2}-\left( 1-a-b-f\right) ^{2}\right) ,
\end{equation}%
and the concurrence is given by%
\begin{equation}
C\left( \varrho \right) =\max \left( 0,C_{1},C_{2}\right)
\end{equation}%
with%
\begin{eqnarray}
C_{1}\left( \varrho \right) &=&d-\sqrt{ab}  \notag \\
C_{2}\left( \varrho \right) &=&c-\sqrt{f\left( 1-a-b-f\right) }
\end{eqnarray}%
The description of the set $\Lambda _{\mathcal{E}_{1}}$ was made
numerically, by generating a very large number of randomly density matrices.
The results (see Fig.\ref{fig6} ) showed that the structure of $\Lambda _{%
\mathcal{E}_{1}}$ is the same of the previous set of density matrices $%
\Lambda _{\mathcal{E}_{0}}$, i.e. the bounds of the the three region $%
\Lambda _{V\_E}$, $\Lambda _{0\_E}$ and $\Lambda _{NV\_E}$ remain unchanged. 

\section{Conclusion}

In this paper, nonlinear inequalities, based on the quadratic Renyi entropy,
were represented on the Entropy-Concurrence plane for the two set of mixed
two-qubit states $\mathcal{E}_{0}$ and $\mathcal{E}_{1}$. and a comparison
was made with respect to CHSH inequalities. The analysis of higher order $%
\left( \alpha >2\right) $ cases shows other interesting properties of
non-linear inequalities and it will be presented successively.


\section{Acknowledgments}

\begin{acknowledgments}
Thanks to Andrea Aiello for useful discussions. The work was supported by
EC-FET project QAP-2005-015848.
\end{acknowledgments}

\end{document}